\documentclass[PAPER, cernpreprint, coverpage=false, atlasdraft=false, texlive=2023, UKenglish, texmf, orcidlogo]{atlasdoc}
 
\usepackage{atlaspackage}
\usepackage{atlasbiblatex}
 
\usepackage{atlasphysics}
 
\graphicspath{{logos/}{figures/}}
 
\usepackage{ANA-HION-2022-14-PAPER-defs}

\AtlasTitle{Comparison of inclusive and photon-tagged jet suppression in 5.02~TeV Pb+Pb collisions with ATLAS}
 
\AtlasAbstract{
Parton energy loss in the quark--gluon plasma (QGP) is studied with a measurement of photon-tagged jet production in 1.7~nb$^{-1}$ of Pb+Pb data and 260~pb$^{-1}$ of \textit{pp} data, both at $\sqrt{s_\mathrm{NN}} = 5.02$~TeV, with the ATLAS detector.
The process \textit{pp} $\to \gamma$+jet+$X$ and its analogue in Pb+Pb collisions is measured in events containing an isolated photon with transverse momentum ($p_\mathrm{T}$) above $50$~GeV and reported as a function of jet $p_\mathrm{T}$.
This selection results in a sample of jets with a steeply falling $p_\mathrm{T}$ distribution that are mostly initiated by the showering of quarks.
The \textit{pp} and Pb+Pb measurements are used to report the nuclear modification factor, $R_\mathrm{AA}$, and the fractional energy loss, $S_\mathrm{loss}$, for photon-tagged jets.
In addition, the results are compared with the analogous ones for inclusive jets, which have a significantly smaller quark-initiated fraction.
The $R_\mathrm{AA}$ and $S_\mathrm{loss}$ values are found to be significantly different between those for photon-tagged jets and inclusive jets, demonstrating that energy loss in the QGP is sensitive to the colour-charge of the initiating parton.
The results are also compared with a variety of theoretical models of colour-charge-dependent energy loss.
}
 
\author{The ATLAS Collaboration}
 
\AtlasRefCode{HION-2022-14}
 
\PreprintIdNumber{CERN-EP-2023-029}

\AtlasJournalRef{Phys. Lett. B 846 (2023) 138154}
\AtlasDOI{10.1016/j.physletb.2023.138154}

\hypersetup{pdftitle={ATLAS document},pdfauthor={The ATLAS Collaboration}}
 
\begin{document}
 
\maketitle
 
\tableofcontents

\section{Introduction}
\label{sec:intro}
 
Ultra-relativistic collisions of heavy nuclei at the Large Hadron Collider (LHC) and the Relativistic Heavy Ion Collider (RHIC) produce a hot, deconfined nuclear medium known as the quark--gluon plasma (QGP). The QGP exhibits interesting emergent phenomena, such as a collective evolution that suggests it is a strongly coupled fluid well described by hydrodynamics~\cite{Romatschke:2017ejr,Heinz:2013th,Elfner:2022iae}. The dense colour field arising from the deconfined colour charges that makes up the QGP is opaque to high-energy quarks and gluons attempting to pass through it. This results in hard-scattered partons suffering energy loss and a modification of their showering processes as they traverse the QGP\@. This phenomenon is known as {\em jet quenching}, and results in a wide variety of experimental signatures -- see Ref.~\cite{Cunqueiro:2021wls} for a recent review.
 
A straightforward and broadly used signature of jet quenching is the suppression of jet production at fixed transverse momentum\footnote{ATLAS uses a right-handed coordinate system with its origin at the nominal interaction point (IP) in the centre of the detector and the $z$-axis along the beam pipe. The $x$-axis points from the IP to the centre of the LHC ring, and the $y$-axis points upward. Cylindrical coordinates $(r,\phi)$ are used in the transverse plane, $\phi$ being the azimuthal angle around the $z$-axis. The pseudorapidity is defined in terms of the polar angle $\theta$ as $\eta=-\ln\tan(\theta/2)$.} (\pt) in Pb+Pb collisions compared to \ppp collisions. This is quantified by the nuclear modification factor, $R_\mathrm{AA}$, which is defined as the ratio of the observed yield in Pb+Pb collisions to the expectation from an equivalent number of nucleon--nucleon (NN) collisions, i.e., without jet quenching effects from the formation of a QGP\@. This expectation is calculated as the cross-section in \ppp collisions, scaled by the mean value of the nuclear thickness function in the corresponding Pb+Pb collisions, $\left<T_\mathrm{AA}\right>$~\cite{Miller:2007ri}. The $R_\mathrm{AA}$ is therefore defined as
\begin{equation}
R_\mathrm{AA} = \left.\frac{1}{N_\mathrm{evt}}\frac{\text{d}^2N^\pbpb}{\text{d}\pt\text{d}\eta} \right/ \left<T_\mathrm{AA}\right> \frac{\text{d}^2\sigma^\ppp}{\text{d}\pt\text{d}\eta},
\label{eq:RAA}
\end{equation}
where $\text{d}^2N^\pbpbNoBlank/\text{d}\pt\text{d}\eta$
is the differential jet yield in $N_\mathrm{evt}$ Pb+Pb events in a given centrality range, 
$\text{d}^2\sigma^\pppNoBlank/\text{d}\pt\text{d}\eta$
is the jet cross-section in \ppp collisions, and $\left<T_\mathrm{AA}\right>$ can be considered as a luminosity of nucleons per Pb+Pb collision. Therefore, the term in the denominator is the expected yield in Pb+Pb collisions in the absence of any nuclear effects.
 
In central Pb+Pb collisions, the nuclei collide head on and create a large and long-lived volume of QGP\@. The developing showers of high-\pt\ partons undergo substantial interactions with the QGP, such that part of their momentum is transfered to large angles relative to the initial parton direction~\cite{CMS-HIN-14-010,HION-2018-03}. Therefore, the total momentum in a fixed-size jet cone is decreased compared to the process with analogous initial kinematics occurring in \ppp collisions, and the jets can be thought of as migrating to lower \pt\ values in Pb+Pb events. Since the jet spectrum is steeply falling with \pt, this results in an $R_\mathrm{AA}$ below unity with a magnitude that depends on the amount of transported energy and the local shape of the spectrum. In central Pb+Pb events at the LHC, the $R_\mathrm{AA}$ for inclusive jets is suppressed by approximately a factor of two at $\pt \approx 100$~\GeV~\cite{HION-2017-10,CMS-HIN-18-014,ALICE:2019qyj}. While the \RAA is expected to be impacted by other effects, such as the modification of parton densities in the nucleus (nPDFs), these are understood to be modest for inclusive jets and thus most of the signal is due to jet energy loss~\cite{Kovarik:2015cma,Eskola:2021nhw,AbdulKhalek:2022fyi}.
 
A key aspect to the theoretical description of jet quenching is its sensitivity to the colour charge of the initiating parton, i.e., whether that parton is a quark or a gluon~\cite{Spousta:2015fca,Mehtar-Tani:2021fud,Takacs:2021bpv,Ke:2018jem,Ke:2018tsh,Ke:2020clc,Kang:2017xnc,Li:2018xuv,Li:2019dre,He:2018xjv,Brewer:2020och,Brewer:2021hmh}. If the jet-medium interaction is predominantly described as proceeding by radiative emission (medium-induced gluon radiation by strong colour charges), quarks and gluons are generally expected to lose energy in proportion to their QCD colour factors for gluon emission of 4/3 and 3, respectively. Thus, gluon-initiated jets are expected to lose significantly more energy than quark-initiated ones. While the developing parton shower eventually contains both quarks and gluons, theoretical models indicate that the charge of the initiating parton should have a significant impact. At LHC energies, inclusive jet production in the region $p_\mathrm{T} < 200$~\GeV\ is dominated by gluon-initiated jets.
 
Several previous measurements have attempted to explore the colour charge dependence of jet suppression, but with additional effects that may complicate its extraction. For example, Ref.~\cite{HION-2017-10} measured jet suppression as a function of jet rapidity, which changes the quark/gluon-initiated jet fraction, but may also sample different regions of the QGP medium~\cite{He:2018xjv}. Refs.~\cite{HION-2018-24,CMS-HIN-12-003} report the suppression of $b$-jets, which have a significantly larger quark-initiated fraction than inclusive jets, but have additional effects from the large mass of $b$-quarks.
 
An alternative strategy, including the one employed in this Letter, is to measure jets produced in association with an isolated photon or other electroweak (EW) boson, for example through Compton scattering ($gq \to q\gamma$). These jets are substantially more likely to be initiated by a quark than inclusive jets at the same \pt. Importantly, the kinematics of the colourless photon or EW boson are not significantly modified by the QGP~\cite{HION-2012-02,CMS-HIN-18-016,HION-2018-25,CMS-HIN-19-006}. Therefore ATLAS has used an isolated photon or $Z$ boson as a way to select partons with a known distribution of initial kinematics before jet quenching~\cite{Wang:1996yh} and to study how the resulting jet~\cite{HION-2018-06} or hadron~\cite{HION-2019-05,HION-2017-08} distributions are modified in particular selections of boson $p_\mathrm{T}$, compared to those in \ppp collisions.
 
This Letter presents a measurement of the process \ppp (or $\textit{NN}$) $\to \gamma$+jet+$X$, as a function of jet \pt. Unlike previous measurements mentioned above (Ref.~\cite{HION-2018-06,HION-2019-05,HION-2017-08}), the results in this paper are not normalized per-photon, but measure the full photon-associated jet production cross-section. The measurement is performed using $260$~pb$^{-1}$ and $1.7$~nb$^{-1}$ of \ppp and Pb+Pb collisions, respectively, at an NN centre-of-mass energy \comHI recorded with the ATLAS detector at the LHC\@. Events are required to have an isolated photon with $p_\mathrm{T}^\gamma > 50$~\GeV\ and $\left|\eta^\gamma\right| < 2.37$ (excluding the region $1.37 < \left|\eta^\gamma\right| < 1.52$). At leading order (LO), the photon isolation requirement predominantly selects direct photons, which are those produced directly in the hard scattering, but also a contribution from fragmentation photons that are radiated in a parton shower after the scattering. All jets with $\left|\eta^\mathrm{jet}\right| < 2.8$ and $\ptj > 50$~\GeV\ in an opposing azimuthal direction to the photon (\dphisevenpieight) are included in the measurement.
This requirement selects a set of jets with a steeply falling \pt\ distribution, with a large quark-initiated fraction.
 
The resulting jet production rates in Pb+Pb and \ppp collisions are used to report $R_\mathrm{AA}$ and the fractional energy loss quantity, $S_\mathrm{loss}$, originally developed by the PHENIX Collaboration at RHIC~\cite{PHENIX:2004vcz,PHENIX:2006wwy,PHENIX:2015vqa} that is conceptually similar to the `pseudo-quantile' described in Ref.~\cite{Brewer:2018dfs}.
For a given amount of energy loss, the particular magnitudes of the $R_\mathrm{AA}$ values are known to depend strongly on the steepness of the \ppp spectrum. The $S_\mathrm{loss}$ formulation is designed as an alternative way to characterize the energy loss while removing this dependence. Schematically, $S_\mathrm{loss}$ is the fractional decrease in \ptj at which the \TAAavg-scaled jet yield in Pb+Pb events reaches the same magnitude as the cross-section in \ppp events at the original \ptj. Quantitatively, for each value of the \ptj in \ppp collisions, $p_\mathrm{T}^\pppNoBlank$, the shift function, $\Delta p_\mathrm{T}(\pt^\pppNoBlank)$, is defined as
\begin{equation}
\Delta\pt = p_\mathrm{T}^\ppp - \pt^\mathrm{Pb+Pb}
\label{eq:deltapt}
\end{equation}
where $p_\mathrm{T}^\mathrm{Pb+Pb}$ is the value for which
\begin{equation}
\frac{1}{\left<T_\mathrm{AA}\right>}\frac{1}{N_\mathrm{evt}}\frac{\text{d}^2N^\pbpb(\pt^\mathrm{Pb+Pb} = \pt^\ppp - \Delta\pt)}{\text{d}\pt^\mathrm{Pb+Pb}\text{d}\eta} =  \frac{\text{d}^2\sigma^\ppp(\pt^\pppNoBlank)}{\text{d}\pt^\ppp\text{d}\eta} \times \left[ 1 + \frac{\text{d}\Delta\pt}{\text{d}\pt^\ppp}\right]
\label{eq:Sloss}
\end{equation}
where the expression in square brackets is the Jacobian term necessary to, e.g., preserve the total number of jets. The fractional energy loss is given by $S_\mathrm{loss}(\ptpp) = \Delta \pt / \ptpp$. It is related to, but not identical to, the average energy lost by jets originating at a given $p_\mathrm{T}$ in \ppp collisions, and is a useful way to characterize the magnitude of energy loss in a way that does not depend on the local shape of the spectrum.
 
The $R_\mathrm{AA}$ and $S_\mathrm{loss}$ results for photon-tagged jets are then compared with the analogous ones for inclusive jets~\cite{HION-2017-10}, whose production in this kinematic range has a significantly smaller quark-initiated fraction. Since the main difference between the jet populations is in their quark and gluon composition, this comparison allows a controlled examination of the impact of the initiating parton's QCD colour charge on jet energy loss.
 
\section{ATLAS detector}
 
The ATLAS detector~\cite{PERF-2007-01} at the LHC is a multipurpose particle detector
with a forward--backward symmetric cylindrical geometry and a near \(4\pi\) coverage in
solid angle.
Its inner tracking detector is surrounded by a thin superconducting solenoid
providing a \SI{2}{\tesla} axial magnetic field, and electromagnetic (EM) and hadron calorimeters.
The inner tracking detector covers the pseudorapidity range \(|\eta| < 2.5\).
It consists of silicon pixel, silicon microstrip, and transition radiation tracking detectors.
Lead/liquid-argon (LAr) sampling calorimeters provide EM energy measurements
with high granularity.
A steel/scintillator-tile hadron calorimeter covers the central pseudorapidity range (\(|\eta| < 1.7\)).
The endcap and forward regions are instrumented with LAr calorimeters
for both the EM and hadronic energy measurements up to \(|\eta| = 4.9\).
A zero-degree calorimeter (ZDC) was situated at $\left|\eta\right| > 8.3$ during Pb+Pb data-taking. It is composed of alternating layers of quartz rods and tungsten plates and is mostly sensitive to spectator neutrons from fragmenting nuclei in Pb+Pb collisions.
 
A two-level trigger system is used to select events~\cite{TRIG-2016-01}.
The first-level trigger is implemented in hardware and uses a subset of the detector information
to accept events at a rate below \SI{100}{\kHz}.
This is followed by a software-based trigger that
reduces the accepted event rate to \SI{1}{\kHz} on average
depending on the data-taking conditions.
An extensive software suite~\cite{ATL-SOFT-PUB-2021-001} is used in data simulation, in the reconstruction and analysis of real
and simulated data, in detector operations, and in the trigger and data acquisition systems of the experiment.
 
\section{Event reconstruction}
 
Events were selected using triggers that required a reconstructed photon with \pt\ above 35~\GeV\ (20~\GeV) in \ppp (\pbpbNoBlank) collisions~\cite{TRIG-2016-01,TRIG-2018-05}. The trigger sampled the full luminosity corresponding to $260$~pb$^{-1}$ of \ppp data in 2017 and $1.7$~nb$^{-1}$ of Pb+Pb data in 2018, and was fully efficient for the photon selection described below. Events are required to satisfy detector and data-quality requirements and, in Pb+Pb collisions, to have a reconstructed vertex.
 
The Pb+Pb event centrality is characterized by the sum of the transverse energy, $\Sigma{E}_\mathrm{T}^\mathrm{FCal}$ in the forward calorimeters, $3.2 < \left|\eta\right| < 4.9$. Events in different ranges of $\Sigma{E}_\mathrm{T}^\mathrm{FCal}$ are associated with an underlying Pb+Pb collision geometry  according to a Monte Carlo (MC) Glauber simulation~\cite{Miller:2007ri,HION-2016-07}. This analysis uses three centrality intervals corresponding to the following fractions of the $\Sigma{E}_\mathrm{T}^\mathrm{FCal}$ distribution in minimum-bias events: 0--10\% (`central' events, with a large nuclear overlap and large $\Sigma{E}_\mathrm{T}^\mathrm{FCal}$ values), 10--30\%, and 30--80\% (`peripheral' events).
 
Photons are reconstructed following the method used previously in Pb+Pb collisions~\cite{HION-2012-02,HION-2018-06,HION-2017-08}, which applies the procedure used in \ppp collisions~\cite{EGAM-2018-01} after an event-by-event estimation and subtraction of the underlying event (UE) contribution to the energy deposited in each calorimeter cell~\cite{HION-2011-02} (described further below). Photon candidates are required to satisfy `tight' shower shape requirements designed to reject photons arising from neutral meson decays and from the start of hadronic showers in the EM calorimeter~\cite{PERF-2017-02}. In \ppp collisions, photons are further required to be isolated by requiring that the sum of the transverse energy in calorimeter cells within $\Delta{R} = 0.3$ (not including the contribution from the photon itself) is less than 3~\GeV. In Pb+Pb collisions, the UE fluctuations within the isolation cone result in a substantial broadening of the isolation $E_\mathrm{T}$ distribution. Thus, in Pb+Pb collisions, the isolation energy requirement depends continuously on the centrality of the event and is chosen so that its efficiency for prompt photons is 90\%, as determined from simulations of photon+jet events overlaid with Pb+Pb minimum-bias data (described in Section~\ref{sec:simulation} below). This upper limit on the isolation energy is approximately 10~\GeV\ in 0--10\% Pb+Pb events, but quickly decreases in more peripheral events and converges to the \ppp value.
 
Jets are reconstructed following the procedure used in Pb+Pb collisions~\cite{HION-2011-02,HION-2017-10}, which is summarized here. Calorimeter cells in all  layers are evaluated at the EM energy scale and regrouped into $\Delta\eta\times\Delta\phi$ = $0.1\times\pi/32$ logical towers, and the anti-$k_t$ algorithm~\cite{Cacciari:2008gp, Cacciari:2011ma} with parameter $R=0.4$ is applied to the towers. After the initial jet-finding, the contribution to the energy deposited in towers by the UE is estimated on an event-by-event basis, allowing for the variation of the UE as a function of $\eta$ and $\phi$ (the latter arising from the global collective flow in Pb+Pb collisions). Information from towers within $\Delta{R} = 0.4$ of jet candidates is excluded to avoid biasing the UE estimate. The kinematics of the tower energies are updated to subtract the estimated UE contribution, and the UE procedure is iterated using a better-defined set of jets to define the exclusion regions. The resulting set of jet kinematics is corrected using $\pt$- and $\eta$-dependent factors, determined from simulation, to account for the response of the calorimeter to jets~\cite{PERF-2012-01}. An additional correction for the absolute response in data is based on {\em in situ} studies of jets recoiling against photons, $Z$ bosons, and jets in other regions of the calorimeter in \ppp collisions~\cite{PERF-2016-04}. This calibration is followed by a `cross-calibration' that relates the jet energy scale (JES) in high-luminosity 13~\TeV\ \ppp collisions~\cite{ATLAS-CONF-2015-016} to the jets reconstructed by the procedure outlined above in the 5.02~\TeV\ data to account for additional differences between the data and simulations.
 
\section{Simulation}
\label{sec:simulation}
 
Samples of MC-simulated events are used to evaluate the performance of the photon and jet reconstruction and to correct the measured distributions for detector effects. The main MC sample corresponding to photon+jet production in \ppp data consists of \PYTHIA8~\cite{Sjostrand:2007gs} events, produced with the A14~\cite{ATL-PHYS-PUB-2014-021} set of tuned parameters (tune) and the NNPDF 2.3 LO~\cite{Ball:2012cx} parton distribution function (PDF) set, including direct and fragmentation contributions. As alternatives, photon+jet events were also produced using two additional generators. The \textsc{Sherpa} 2.2.4~\cite{Bothmann:2019yzt,Siegert:2016bre} generator was run at next-to-leading order (NLO) with the NNPDF 3.0 NNLO~\cite{Ball:2014uwa} PDF set to produce a sample of events containing a photon plus up to three other partons. The \textsc{Herwig} 7.2~\cite{Bellm:2015jjp} generator was run at leading order with the MMHT2014lo~\cite{Harland-Lang:2014zoa} PDF set, with separate samples produced for direct and fragmentation photons. The three sets of events were simulated~\cite{ATL-SOFT-PUB-2021-001} using a \GEANT~\cite{Agostinelli:2002hh} description of the ATLAS detector and were digitized and reconstructed in a manner identical to that of the data. The generator-level final state photons in the MC samples are required to be isolated by requiring that the sum of the transverse energy of all the final state particles, excluding the photon itself, within a $\Delta{R} = 0.4$ cone is less than 5~\GeV.
 
\begin{figure}[hbtp!]
\begin{center}
\includegraphics[width=0.48\textwidth]{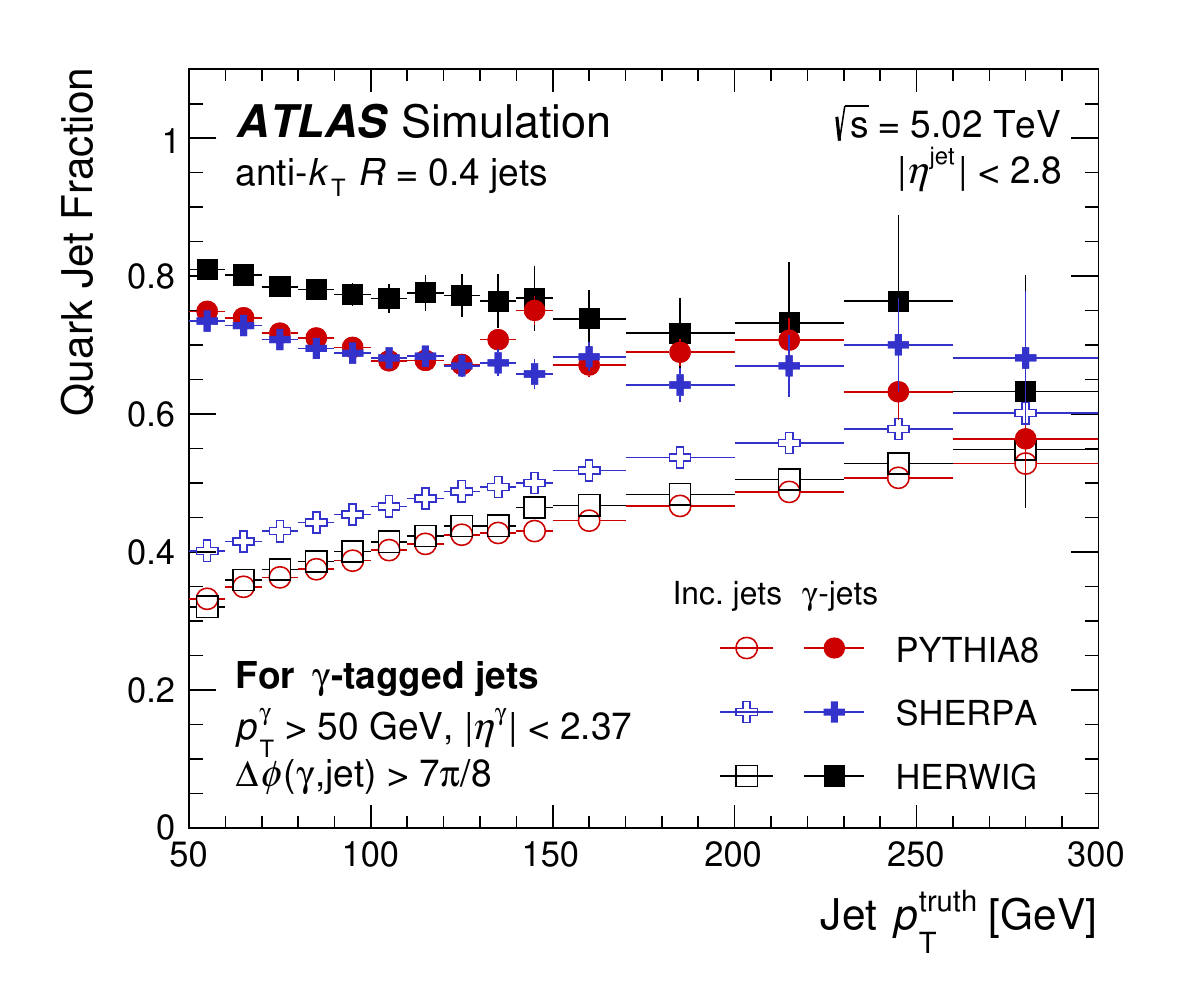}
\caption{
Fraction of photon-tagged jets (filled markers) and inclusive jets (open markers) initiated by a quark, as a function of generator-level \ptj, in the \PYTHIA8 (circles), \HERWIG (squares), and \SHERPA (crosses) event generators. The vertical bars associated with symbols indicate the statistical uncertainties.
}
\label{fig:res_quarkFraction}
\end{center}
\end{figure}

The fraction of quark-initiated jets, as defined in simulation in Ref.~\cite{PERF-2013-02}, is estimated by using three different MC generators (\PYTHIA8, \HERWIG\ and \SHERPA) for the photon-tagged jets, and is compared with that for inclusive jets~\cite{HION-2017-10} in Figure~\ref{fig:res_quarkFraction}.
The generators predict that 75--80\% of all photon-tagged jets at $p_\mathrm{T}^\mathrm{jet} = 50$~\GeV\ are initiated by quarks, while this is true for only 30--40\% of inclusive jets at the same \ptj. At higher \ptj, the quark-initiated fractions for photon-tagged and inclusive jets slowly fall and rise, respectively, reaching 50--60\% for both samples at 300~\GeV. Thus, according to the MC generators, these two samples contain significantly different quark-initiated jet fractions with \ptj $\lesssim$ 200~\GeV.
 
To simulate photon+jet events in Pb+Pb data, the events described above were overlaid at the detector-hit level with a sample of Pb+Pb data events recorded with minimum-bias and central-event triggers. The combination of the simulated and data event was then reconstructed as a single event. These `Pb+Pb data overlay' events are re-weighted to match the  observed $\Sigma{E}_\mathrm{T}^\mathrm{FCal}$ distribution for photon+jet events in Pb+Pb data. In this way, the features of the Pb+Pb UE in the simulated samples which are uncorrelated with the photon+jet process, such as the flow and transverse energy distributions, are identical to those in real minimum-bias Pb+Pb events.
 
Finally, to evaluate the possible impact of nuclear effects, such as the modification of PDFs on the measurement, samples of generator-level \PYTHIA8 events were produced for photon+jet and inclusive jet events, again including both direct and fragmentation photons and the generator-level isolation requirement.
For both of these processes, separate samples were generated for \textit{pp}, proton--neutron (\textit{pn}), and neutron--neutron (\textit{nn}) events, and the cross-section in simulated Pb+Pb events was constructed via a weighted sum $\left(Z^2\sigma^{\pppNoBlank} + 2Z(A-Z)\sigma^{\textit{pn}}+(A-Z)^2\sigma^{\textit{nn}}\right) / A^2$, where $A$ and $Z$ are the mass and atomic number of Pb, respectively. In a separate procedure, the cross-section in the \ppp samples was evaluated after being weighted on an event-by-event basis with the central values of the EPPS16 nPDF set~\cite{Eskola:2016oht}, at NLO and configured for the lead nucleus with the grid file \textsc{EPPS16NLOR\_208}.
 
\section{Analysis}
 
The signal definition for this measurement is $R=0.4$ jets with $\ptj > 50$~\GeV\ that are \dphisevenpieight, i.e. $|\Delta \phi_{\mathrm{\gamma,jet}} - \pi|\mathrm{\ < \pi/8}$, from a $\ptg > 50$~\GeV\ isolated photon, with all candidate jets in a given event included in the measurement. The two-dimensional yield $(\pt^\gamma,\pt^\mathrm{jet})$ is constructed for photons and their associated jets, but using thresholds of 40~\GeV\ on the photon and jet \pt, to allow for the correction of bin migration effects (discussed below).
 
\begin{figure}[t!]
\begin{center}
\includegraphics[width=0.90\textwidth]{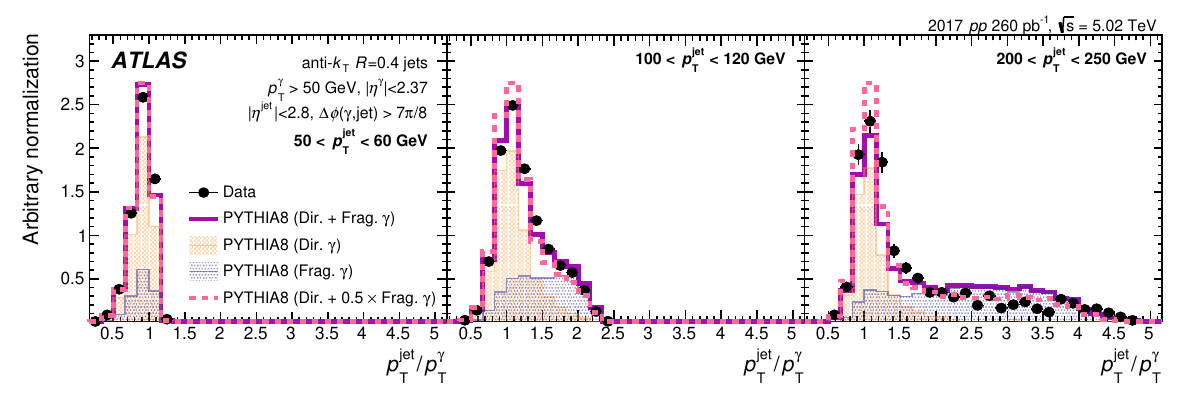}
\caption{
Reconstructed-level \ptj/\ptg distributions for different \ptj bins: (left panel) 50 $<$ \ptj $<$ 60~\GeV\, (middle panel) 100 $<$ \ptj $<$ 120~\GeV\, and (right panel) 200 $<$ \ptj $<$ 250~\GeV\ for \ptg $>$ 50~\GeV.
The data (filled circles) is compared with reconstructed-level \PYTHIA8 (solid line), which is normalized to the data. The shaded histograms show the breakdown of \PYTHIA8 contributions from photon-production processes: direct (checkered hatching) and fragmentation (dashed hatching) photons. The dashed-line histogram represents \PYTHIA8 events, including all direct photon events with the fragmentation photon events scaled down by a factor of two.
}
\label{fig:paper_xjg}
\end{center}
\end{figure}

Figure~\ref{fig:paper_xjg} shows the signal \ptj/\ptg distributions in \ppp data at the reconstructed-level (i.e., without any of the corrections for photon purity, efficiency, and unfolding described below), compared with the same in simulated \PYTHIA8 events.
The contributions from direct and fragmentation photons in \PYTHIA8 are shown separately as shaded histograms, with the former contribution peaking near unity due to the back-to-back kinematics, and the latter distribution extending to large \ptj/\ptg values.
At the lowest \ptj bin of $50 < \ptj < 60$~\GeV, the \ptj/\ptg distribution in data has no entries above 1.2 because of the kinematic selection on the photons ($\ptg > 50$~\GeV), and thus the comparison with simulation suggests that direct photons are dominant. However, at high \ptj values (e.g., in the right most panel), there is a growing contribution from fragmentation photons, which may contribute to the decreasing quark-initiated jet fraction in Figure~\ref{fig:res_quarkFraction}.
 
Notably, \PYTHIA8 does not precisely match the \ptj/\ptg distribution in data, in particular over-estimating the relative magnitude of the fragmentation photon contribution. A similar conclusion was reached in the study of photon+jet events in \ppp collisions at 7~\TeV~\cite{STDM-2012-18}, where \PYTHIA8 better describes the data after an increased (decreased) weighting of the direct (fragmentation) contributions in that generator. While this exercise is not repeated in this measurement, the dashed line in Figure~\ref{fig:paper_xjg} indicates how de-weighting the fragmentation photon contribution in \PYTHIA8 by, e.g., a factor of two would modify the jet $p_\mathrm{T}$ distribution in that generator. Therefore this study highlights the need for the \ppp baseline in theoretical calculations of jet quenching to properly model the relative direct and fragmentation photon contributions in photon+jet processes.
 
The initial $(\pt^\gamma,\pt^\mathrm{jet})$ yield in Pb+Pb collisions contains jets that do not arise from the same hard scattering as the photon, but rather from an unrelated NN scattering, or from jets that are reconstructed from the localized fluctuations of the UE\@. This combinatoric contribution is estimated through a `mixing' technique in which high-\pt\ photons in data are correlated with jets in minimum-bias Pb+Pb events that match the overall properties of the original, photon-containing event. These matched properties include the $\Sigma{E}_\mathrm{T}^\mathrm{FCal}$ and flow plane angle. The resulting combinatoric jet contribution is observed to be flat in $\Delta\phi$, as expected for unrelated pairs. For the lowest \ptj values in the most central events, the background contribution is approximately half of the total yield, but this fraction falls very rapidly with increasing \ptj or in less central events. The contribution is statistically subtracted from the initial yields.
 
Even after the photon identification and isolation conditions above are applied to data, the selected photons still include a considerable contribution from backgrounds, dominantly from neutral hadron decays (e.g., $\pi^{0},\eta\to \gamma\gamma$). These decay photons may be reconstructed as a single cluster that satisfies the `tight' identification and the isolation conditions. Thus, the photon-associated jet yields contain a contribution from, e.g., $\pi^0$-associated jet yields. To correct for this, the purity of prompt, isolated photons in the selected data sample is determined by using a data-driven, double-sideband method widely used in ATLAS photon measurements~\cite{STDM-2014-09,STDM-2016-08,STDM-2012-16,STDM-2015-11}, separately for each selection in event centrality and \ptg. The purity has a minimum of $\approx75$\% in central Pb+Pb events at the lowest \ptg values, but then increases rapidly with \ptg and in more peripheral Pb+Pb or \ppp events to a plateau of $\approx95$\%. The shape of the $\pt^\mathrm{jet}$ contribution from this background is determined by performing the same analysis but using an inverted signal selection on the photon. This selection requires the photon to still be isolated, but fail to satisfy several shower shape requirements in a way that is designed to greatly enhance the neutral hadron background. Finally, the background level is scaled according to the purity in each $\ptg$ and centrality selection, and statistically subtracted from the yields.
 
To correct for the bin-to-bin migration in the $\ptg$ and $\ptj$ distributions arising from the finite detector resolution and residual defects in the JES, a two-dimensional unfolding procedure on the background-subtracted $(\pt^\gamma,\pt^\mathrm{jet})$ yields is used. The \PYTHIA8 simulation samples are used to generate independent response matrices for \ppp events and for each centrality range in Pb+Pb events, after reweighting the \ptj\ distributions in simulation to match those measured in data. The iterative Bayesian method~\cite{DAgostini:1994fjx} is used with the \textsc{RooUnfold} software package~\cite{Adye:2011gm}. The number of iterations used in the unfolding is determined by minimizing the sum in quadrature of the total statistical uncertainty and the differences in the unfolded distribution between consecutive iterations. This number is two or three depending on the event centrality. The unfolding procedure also accounts for the finite reconstruction and selection efficiency for photons, which is $\approx70$\% at low-\ptg in central Pb+Pb events, but rises rapidly with \ptg and in more peripheral events to a plateau of $\approx85$\%, and for a small inefficiency for jets at low \ptj. When tested in simulation, this unfolding procedure leads to a recovery of the original generator-level distribution within the statistical uncertainties of the test sample.
 
\section{Systematic uncertainties}
\label{sec:sysUncertainty}
 
The main sources of systematic uncertainty in this measurement are those associated with the photon, jet, and unfolding components. For most of the sources described below, the entire analysis is repeated with a given variation, and the change in the results is taken as the corresponding uncertainty. These individual uncertainties are treated as independent and added in quadrature to quantify the full uncertainties.
 
The photon measurement includes several uncertainty components. First, the reconstructed energy of photons in simulation is varied according to the uncertainties in the photon energy scale and resolution~\cite{PERF-2017-03}. Second, the reconstructed shower shape variables used to identify photons are varied in simulation~\cite{PERF-2017-02}. Third, the isolation and identification sideband boundaries used in purity determination are varied in a manner similar to that in Refs.~\cite{HION-2018-06,HION-2017-08}. Fourth, the difference between using the nominal purity values and the results of a smooth fit to those values is considered. Finally, the reconstruction-level isolation energy requirement is varied such that isolation efficiency for signal photons is 85\% and 95\%, instead of the nominal 90\%. These variations result in different estimates of the photon purity, and thus test the stability of the extracted yield to any potentially imperfect description of photon isolation energy distributions in simulations. The uncertainty in the yields from all these sources is typically 3--6\% in \ppp collisions (4--15\% in central Pb+Pb collisions), rising with jet $p_\mathrm{T}$.
 
For the jet-related uncertainties, the reconstructed jet energy in simulation is varied according to the uncertainties in the JES and jet energy resolution (JER). As in other Run~2 heavy-ion jet measurements~\cite{HION-2018-06,HION-2017-08,HION-2017-10,HION-2020-09}, the JES uncertainties have four main components. First, a centrality-independent baseline component determined from {\em in situ} studies of the calorimeter response to jets reconstructed following the procedure used in 13~\TeV\ \ppp collisions~\cite{PERF-2011-03,PERF-2016-04}. Second, a centrality-independent component accounting for the relative energy scale difference between the heavy-ion jet reconstruction in this analysis and that used for 13~\TeV\ \ppp collisions~\cite{ATLAS-CONF-2015-016}. Third, a component that accounts for potential inaccuracies in the relative abundances of jets initiated by quarks and gluons, and of their different calorimetric response, in simulation. This uncertainty was evaluated by using the flavour fractions and flavour-dependent response in the \textsc{Herwig}, instead of \PYTHIA8, simulation samples. Finally, a centrality-dependent component accounting for a different structure and possibly a different detector response of jets in Pb+Pb collisions that is not modelled in simulation. This uncertainty is determined by the method used for 2015 and 2011 data~\cite{ATLAS-CONF-2015-016} that compares the calorimeter \ptj with the \pt sum of the charged particles in the jets in data and simulation. For the JER uncertainty, the reconstructed \ptj\ in simulation is smeared by a factor evaluated using an {\em in situ} technique in $13$~\TeV\ \ppp data~\cite{PERF-2011-04,PERF-2014-02}, and by an additional contribution to account for the differences between the heavy-ion jet reconstruction and that in the $13$~\TeV\ \ppp data. The JES and JER uncertainties in the jet yields are typically 3--7\% in \ppp collisions, rising slowly with jet $p_\mathrm{T}$, and are modestly higher in Pb+Pb collisions due to the final uncertainty source described above.
 
Two uncertainties associated with the unfolding procedure are evaluated. First, the impact of a different prior in the response matrices was determined by not applying the reweighting factors to account for the difference in the distributions between data and simulation. These were at most 5\% at low \ptj, decreasing to 1\% at high \ptj. Second, a resampling study is used to determine the impact on the results from the limited size of the simulated samples. These are included as part of the statistical uncertainties, but they are typically much smaller than the statistical uncertainties in data.
 
The mixed event technique was tested in the simulation samples, where the combinatoric contribution is exactly known. Any ``non-closure'' in the procedure (i.e. failure to fully subtract the combinatoric contribution) is considered as a source of uncertainty. Finally, there are uncertainties in the overall normalization of the measurements. For the \ppp cross-section, these arise from the luminosity of the \ppp data and are estimated to be 1.6\% using the beam separation scan analysis methods similar to that in Ref.~\cite{ATLAS:2022hro}. 
For the $1/\left<T_\mathrm{AA}\right>$-scaled yields in Pb+Pb collisions, the uncertainties are determined by adjusting the parameters in the Glauber analysis~\cite{Miller:2007ri,HION-2016-07}, and vary from 0.5\% to 2.8\% in central to peripheral collisions, respectively.
 
Uncertainty sources that are correlated between Pb+Pb and \ppp collisions, which include most of the jet- and photon-related uncertainties, typically cancel out to a large degree in \raa. The most significant uncorrelated uncertainties are the centrality-dependent JES and unfolding ones.
 
For both the cross-section and \RAA measurements, the unfolding (photon purity) uncertainties are dominant at \ptj $<$ 80~\GeV\ for the 0--10\% and 10--30\% centrality intervals (30--80\% centrality interval and in \ppp collisions). At 80 $<$ \ptj $<$ 200~\GeV, the JES, JER and photon purity uncertainties are dominant in all centrality bins and in \ppp collisions. The photon isolation uncertainties are dominant at \ptj $>$ 200~\GeV\ in all centrality bins and in \ppp collisions. In comparisons of the value of \raa reported in this paper to that measured for inclusive jets~\cite{HION-2017-10}, the uncertainties in the two measurements are treated as uncorrelated. For the $S_\mathrm{loss}$ analysis, these uncertainties are propagated as part of the $S_\mathrm{loss}$ determination procedure, described below in Section~\ref{ssect:Sloss}.
 
\section{Results}

\begin{figure}[t!]
\begin{center}
\includegraphics[width=0.48\textwidth]{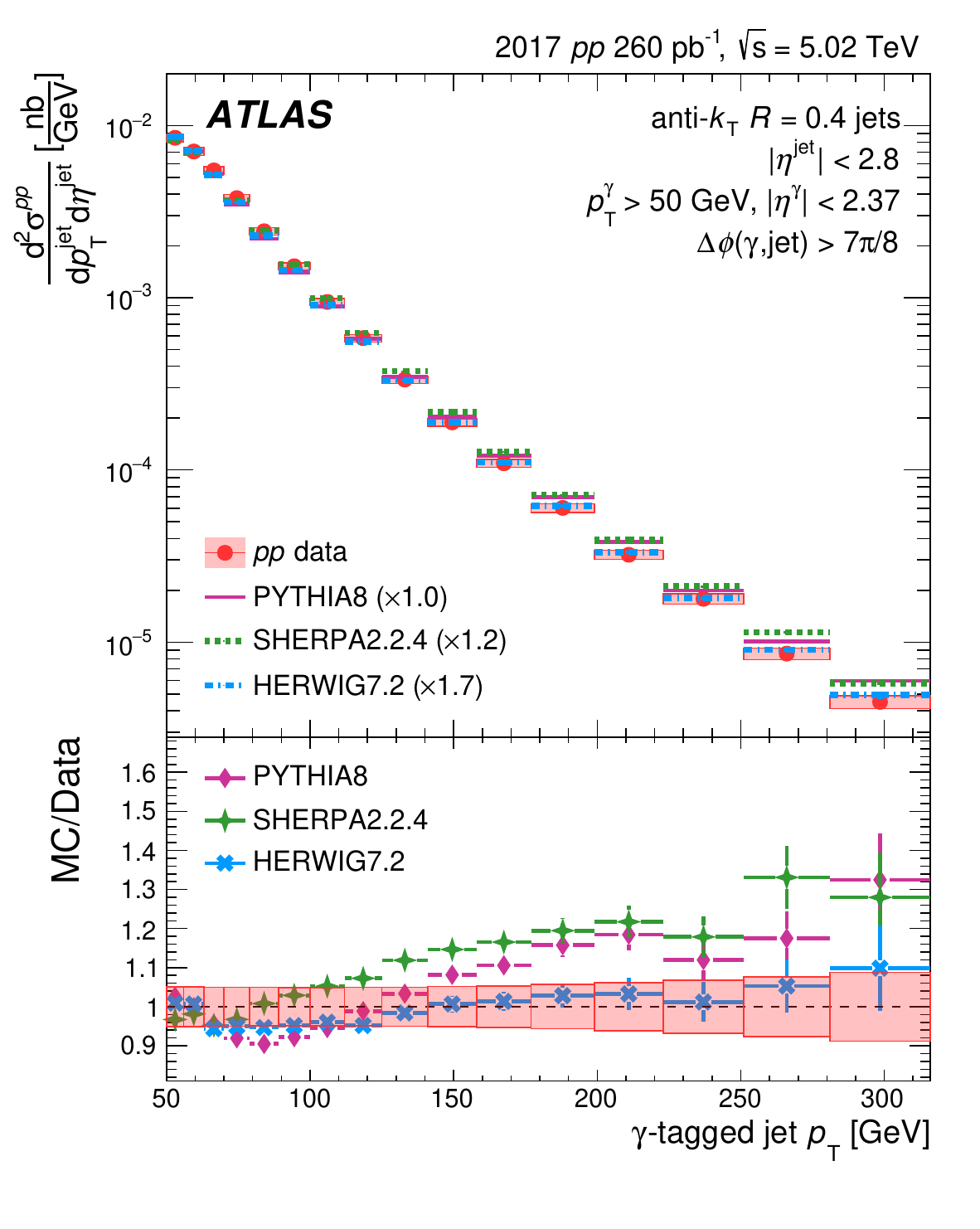}
\caption{
Top panel: The differential cross-section of photon-tagged jets as a function of \ptj in \ppp data, compared with that in \PYTHIA8 (solid line), \SHERPA2.2.4 (dotted line) and \HERWIG7.2 (dash-dotted line) MC samples.
The statistical uncertainties in the data are small and hidden by the symbols, and are drawn as vertical bars for the MC samples. The total systematic uncertainties in the data are shown as boxes in each \ptj bin. The MC distributions are normalized using the factors shown in parentheses to have the same total cross-sections as the data.
Bottom panel: The ratio of cross-sections from different MC generators to the data.
}
\label{fig:xSec_data_MC_comp}
\end{center}
\end{figure}
 
Figure~\ref{fig:xSec_data_MC_comp} shows the measured cross-section for photon-tagged jet production in \ppp collisions, compared with the same quantity in the \PYTHIA8, \HERWIG, and \SHERPA event generators. The distributions of the generators are normalized to have the same total cross-sections as the data.
The data is best described by \HERWIG, which has a shape compatible with the data within its uncertainties over the entire measured \ptj range. \PYTHIA8 and \SHERPA are compatible with the data in the low \ptj region ($\ptj < 100$~\GeV) but have a higher relative cross-section than the data at higher \ptj. The level of agreement between the MC generators and the data has a similar magnitude and $p_\mathrm{T}$ dependence as that observed in previous measurements in \ppp collisions at 7~\TeV~\cite{STDM-2012-18}.

\begin{figure}[hbtp!]
\begin{center}
\includegraphics[width=0.48\textwidth]{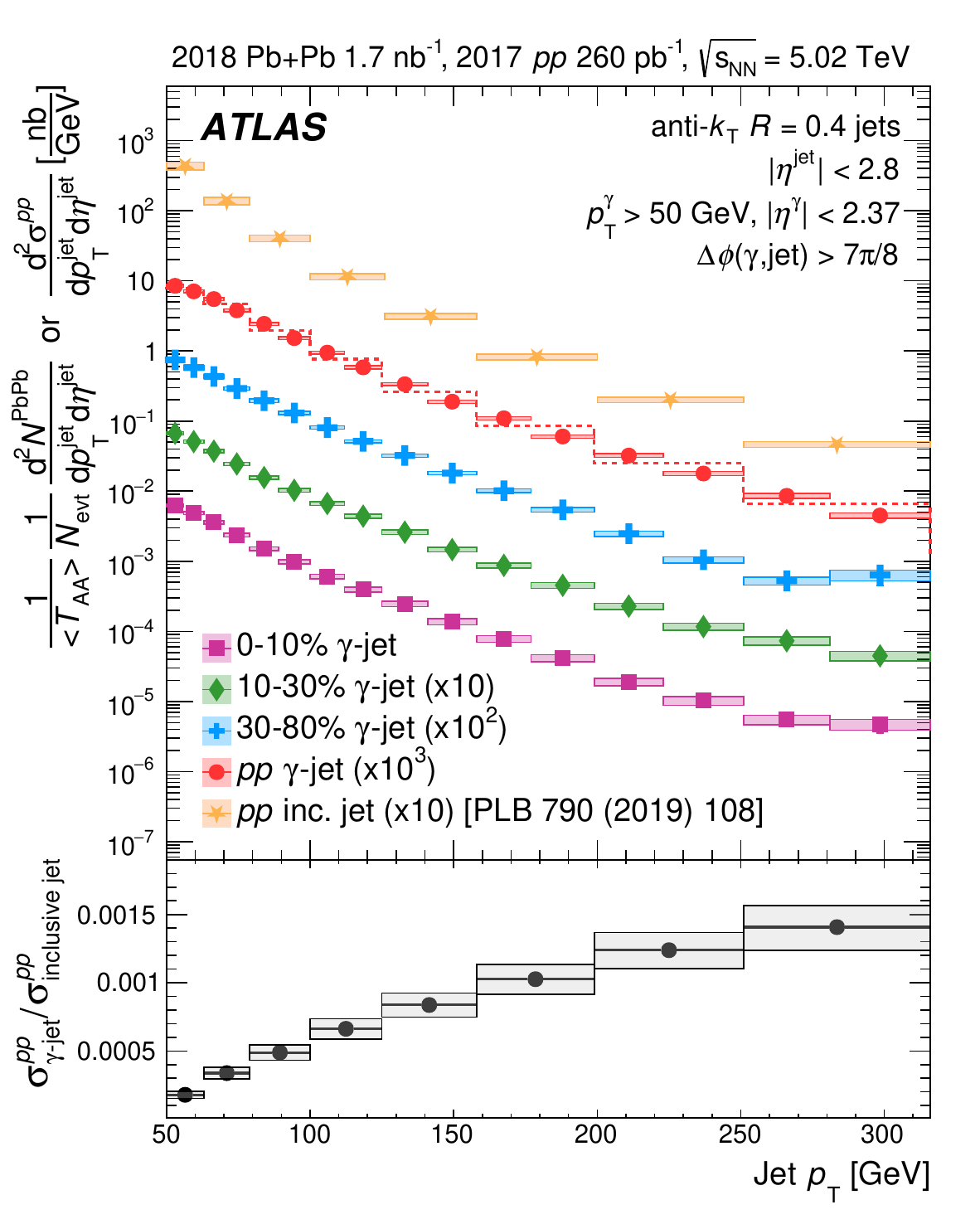}
\caption{
Top panel: The yields of photon-tagged jets as a function of \ptj in \pbpb events for 0--10\% (squares), 10--30\% (diamonds) and 30--80\% (crosses) centrality bins and the differential cross-section in \ppp events (circles). The spectra are scaled by the factors shown in the legend for clarity. The inclusive jet cross-section in \ppp collisions~\cite{HION-2017-10} (stars) is shown for comparison.
The statistical uncertainties are small and hidden by the symbols. The total systematic uncertainties are shown as boxes in each \ptj bin. The photon-tagged jet \ppp data is also re-binned to match the binning of inclusive jet data (shown as dotted line). Bottom panel: The ratio of cross-sections between photon-tagged jets and inclusive jets in \ppp collisions. The boxes associated with the data points represent the sum in quadrature of the systematic uncertainties for photon-tagged jets and inclusive jets.
}
\label{fig:xSec}
\end{center}
\end{figure}

Figure~\ref{fig:xSec} shows the \TAAavg-scaled photon-tagged jet yields for different centrality bins in \pbpb collisions and the cross-section in \ppp collisions. The ratio of cross-sections for photon-tagged jets to that for inclusive jets in \ppp collisions is shown in the bottom panel. Both the \ppp inclusive jet and photon-tagged jet cross-sections are steeply falling as a function of \ptj, but the photon-tagged jet cross-section has a less steep spectrum, i.e., it decreases more slowly with \ptj.  As described above, the \RAA depends on the convolution of the energy loss due to jet quenching with the slope of the \ptj spectrum and this must be taken into account when comparing results between inclusive and photon-tagged jets.
 
\begin{figure}[hbt!]
\begin{center}
\includegraphics[width=0.58\textwidth]{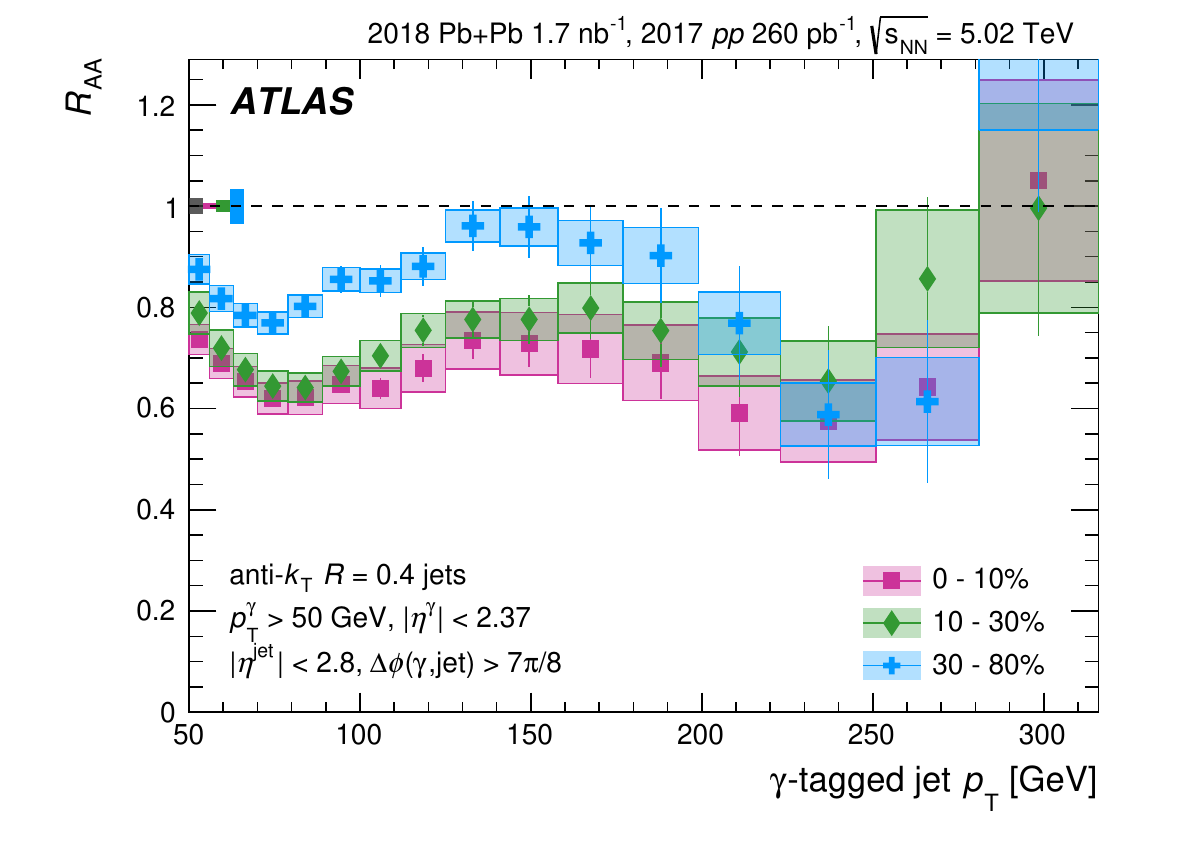}
\caption{
The \RAA of photon-tagged jets as a function of \ptj for 0--10\%, 10--30\%, and 30--80\% centrality intervals.
The vertical bars associated with symbols indicate the statistical uncertainties.
The total systematic uncertainties are shown as boxes in each \ptj bin.
The shaded bars on the left of the axis at $\RAA = 1$ indicate the $p_\mathrm{T}$-independent uncertainties associated with the luminosity in \ppp collisions and \TAAavg for 0--10\%, 10--30\%, and 30--80\% Pb+Pb collisions, respectively. The highest \ptj data point in the 30--80\% centrality interval is 1.42 $\pm$ 0.43 (stat.) $\pm$ 0.25 (syst.) and extends off the vertical scale.
}
\label{fig:Raa_allcentbins}
\end{center}
\end{figure}
 
The \RAA values of photon-tagged jets are computed according to Eq.~(\ref{eq:RAA}) above, and are shown in Figure~\ref{fig:Raa_allcentbins} as a function of \ptj in different centrality intervals. In 0--10\% \pbpb collisions, the \RAA for photon-tagged jets is suppressed below unity, as expected from jet energy loss, and ranges between 0.60--0.75 depending on the jet $p_\mathrm{T}$. Below 70~\GeV, as the jet $p_\mathrm{T}$ decreases, the \RAA values systematically increase. As the \RAA depends not only on the energy loss but on the local shape of the initial spectrum, this increase may be related to the flattening of the spectrum near $p_\mathrm{T}^\mathrm{jet} = 50$~\GeV, which is caused by the kinematic selection $p_\mathrm{T}^\gamma > 50$~\GeV.
In the region $p_\mathrm{T}^\mathrm{jet} < 200$~\GeV, the \RAA is found to be larger in the 30--80\% \pbpb collisions than in the 0--10\% \pbpb collisions, indicating more suppression in central \pbpb collisions as expected due to a larger jet quenching effect in collisions with a larger volume and higher temperature QGP.
 
\FloatBarrier
 
\section{Discussion}
 
\begin{figure}[h!]
\begin{center}
\includegraphics[width=0.58\textwidth]{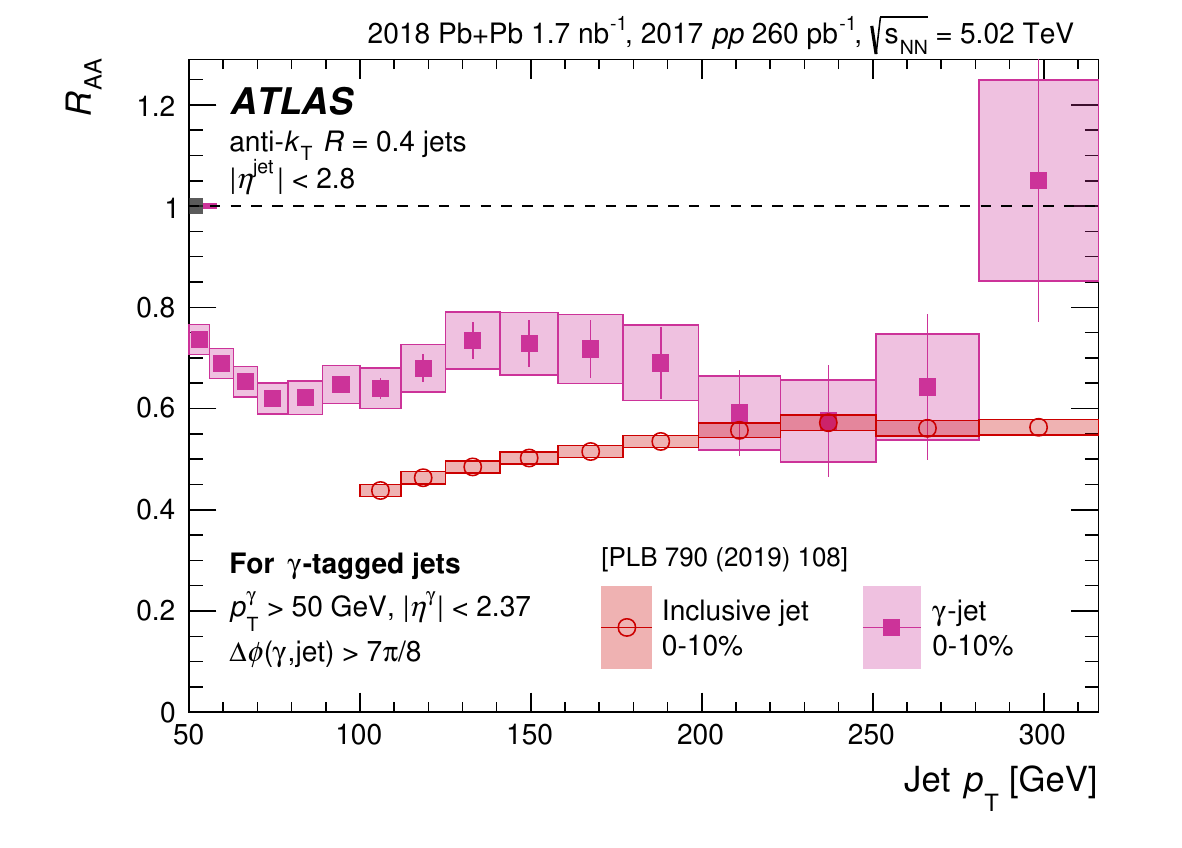}
\caption{
The \RAA of photon-tagged jets (filled squares) as a function of \ptj for 0--10\% \pbpb events are overlaid with that of inclusive jets~\cite{HION-2017-10} (open circles) in the same centrality range for comparison.
The vertical bars associated with symbols indicate the statistical uncertainties.
The total systematic uncertainties are shown as boxes in each \ptj bin.
The shaded bars on the left of the axis at $\RAA = 1$ indicate the $p_\mathrm{T}$-independent uncertainties associated with the luminosity in \ppp collisions and \TAAavg for 0--10\% Pb+Pb collisions, respectively.
}
\label{fig:Raa_0to10}
\end{center}
\end{figure}

Figure~\ref{fig:Raa_0to10} compares the photon-tagged jet \RAA results to the previously published ATLAS inclusive jet results~\cite{HION-2017-10}.
The \RAA of photon-tagged jets is significantly higher than the corresponding values for inclusive jets for $\ptj < 200$~\GeV. For $\ptj > 200$~\GeV, the statistical and systematic uncertainties in the photon-tagged jet results are larger and the two sets of \RAA values become compatible.
 
A primary goal of this measurement is to isolate the effect of colour charge on jet quenching.  Indeed, in the range of \ptj where the quark-initiated fraction is significantly higher in photon-tagged jets (see Figure~\ref{fig:res_quarkFraction}), the \RAA is significantly higher than that for inclusive jets. However, the \RAA is known to depend on the shape of the initial production spectrum with, e.g., a steeper spectrum resulting in a lower $R_\mathrm{AA}$ for the same magnitude of energy loss. Indeed, Figure~\ref{fig:xSec} shows that although the jet $p_\mathrm{T}$ spectra for photon-tagged and inclusive jets are both steeply falling, the latter is systematically steeper than the former.
Thus, it is important for theoretical calculations attempting to describe the \RAA results to first correctly describe the photon-tagged and inclusive jet cross-sections in \ppp collisions, i.e., before applying any jet quenching.
 
\subsection{Fractional energy loss analysis}
\label{ssect:Sloss}
 
An alternative way to characterize the energy loss with a greatly reduced sensitivity to the spectral shape is through the fractional energy loss quantity, $S_\mathrm{loss}$, introduced in Section~\ref{sec:intro}.
 
To determine $S_\mathrm{loss}$ for the photon-tagged jet case, the distributions in \ppp and Pb+Pb collisions are fit using the `extended power law' function introduced in Ref.~\cite{Spousta:2015fca}, $f(p_\mathrm{T}) = A (p_\mathrm{T,0} / p_\mathrm{T})^{n+\beta\log(p_\mathrm{T} / p_\mathrm{T,0}) }$, in the region $p_\mathrm{T} > 100$~\GeV. An initial estimate of $\Delta\pt$ in Eq.~(\ref{eq:deltapt}) is performed by first assuming that the Jacobian term in Eq.~\ref{eq:Sloss}, $\left(1 + \text{d}\Delta\pt/\text{d}\pt^\pppNoBlank\right)$, is unity, i.e. $\text{d}\Delta\pt/\text{d}\pt^\ppp = 0$, and determining $\Delta\pt(\pt^\pppNoBlank)$ from the fitted functions. This estimate is then iteratively improved by applying the Jacobian factor to the \ppp spectrum and repeating the procedure to obtain an updated estimate of $\Delta\pt$. To determine the systematic uncertainty in $\Delta\pt$, and thus $S_\mathrm{loss}$, the procedure is performed separately under each of the systematic variations detailed in Section~\ref{sec:sysUncertainty}, with the variations from sources that are correlated between \ppp and Pb+Pb applied to both distributions simultaneously. An additional uncertainty is assigned to account for the sensitivity of the extracted $S_\mathrm{loss}$ values to the choice of fit range, which is sub-dominant to the other sources described in Section~\ref{sec:sysUncertainty}.
 
To determine $S_\mathrm{loss}$ for the inclusive jet case, this procedure is repeated with two modifications. First, to provide a better description of the data, the fit function for the inclusive jet distributions includes an additional term in the exponent that is linear in $p_\mathrm{T}$. Second, an alternative procedure is used to account for the correlated uncertainties between the \ppp and Pb+Pb distributions. The $\left<T_\mathrm{AA}\right>$-scaled Pb+Pb yields are re-calculated by taking the $R_\mathrm{AA}$ values (including their uncertainties, which account for the correlation between \ppp and Pb+Pb) and multiplying them by the central values of the \ppp cross-section. Then, the uncertainties in $\Delta\pt$, and thus in $S_\mathrm{loss}$, are calculated by propagating the uncertainty in the determined value of $\pt^\mathrm{Pb+Pb}$ using Eq.~(\ref{eq:deltapt}).
 
\begin{figure}[t!]
\begin{center}
\includegraphics[width=0.65\textwidth]{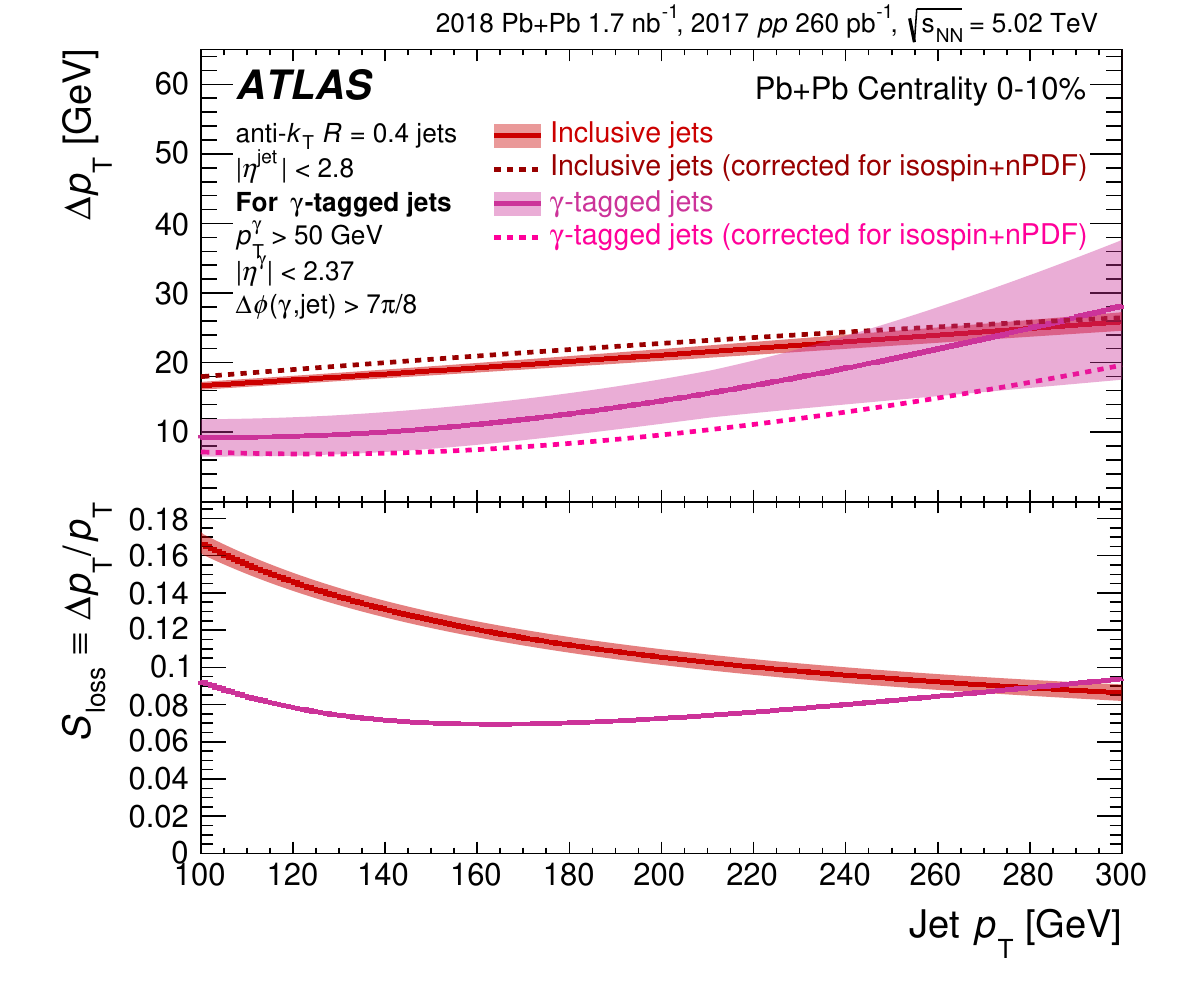}
\caption{
Top panel: The energy loss $\Delta p_\mathrm{T}$
as a function of \ptj for photon-tagged jets (lower bands) and inclusive jets (upper bands) for the 0--10\% centrality interval. The bands around the solid lines indicate the systematic uncertainties. The dashed lines show the updated estimate of $\Delta p_\mathrm{T}$ when the data are corrected for isospin and nPDF effects (see text).
Bottom panel: The fractional energy loss $S_\mathrm{loss}$.
}
\label{fig:eloss_incJet_phoJet}
\end{center}
\end{figure}
 
The extracted $\Delta p_\mathrm{T}$ and $S_\mathrm{loss}$ values are shown in Figure~\ref{fig:eloss_incJet_phoJet} for photon-tagged jets and inclusive jets for the 0--10\% centrality interval. For photon-tagged jets, $\Delta p_\mathrm{T}$ ranges from 10--30~\GeV, and $S_\mathrm{loss}$ from 0.07--0.10. For both samples, $\Delta p_\mathrm{T}$ increases with jet \pt. In the inclusive jet case, this increase is slower than the jet \pt, resulting in $S_\mathrm{loss}$ values that instead decrease systematically with increasing $p_\mathrm{T}$. For the photon-tagged jet case, the $S_\mathrm{loss}$ values are approximately constant within uncertainties over this \ptj range. In the region $100 < p_\mathrm{T}^\mathrm{jet} \lesssim 200$~\GeV, the $S_\mathrm{loss}$ values for photon-tagged jets are significantly smaller than those for inclusive jets, again suggesting a significant colour-charge dependence to jet energy loss. At higher \ptj, the two $S_\mathrm{loss}$ curves are compatible within uncertainties, potentially due to the quark fractions of the two samples becoming more similar in this \pt region (Figure~\ref{fig:res_quarkFraction}).
 
Importantly, $S_\mathrm{loss}(\pt)$ should not be interpreted as the fraction of the energy lost in the QGP for jets that emerge with the given \pt in Pb+Pb collisions. As detailed in Refs.~\cite{PHENIX:2006wwy,PHENIX:2015vqa}, this extracted value is smaller than the true average energy loss. This is due to the steeply falling \pt spectrum and jet-to-jet fluctuations in the energy loss, which result in the fact that jets observed in Pb+Pb at a given $p_\mathrm{T}$ are more likely to be those with smaller than average energy loss. Nevertheless, the procedure above is clearly defined and is a useful way to quantify the difference in the magnitude of energy loss between different scenarios.

Even though the determination of $S_\mathrm{loss}$ is not strongly sensitive to the initial \ptj shape in \ppp collisions, there are other effects that modify the jet spectra in Pb+Pb collisions compared to those in \ppp collisions, which do not arise from energy loss but may impact the extracted $S_\mathrm{loss}$ values. These include effects originating from isospin (i.e., the different up- and down-quark composition of the nucleus compared to the proton, which decreases the rate of processes such as photon+jet production, as previously observed in $p$+Pb collisions~\cite{HION-2018-05}) and the modification of the PDFs in nuclei compared to those in free nucleons.

\begin{figure}[t!]
\begin{center}
\includegraphics[width=0.58\textwidth]{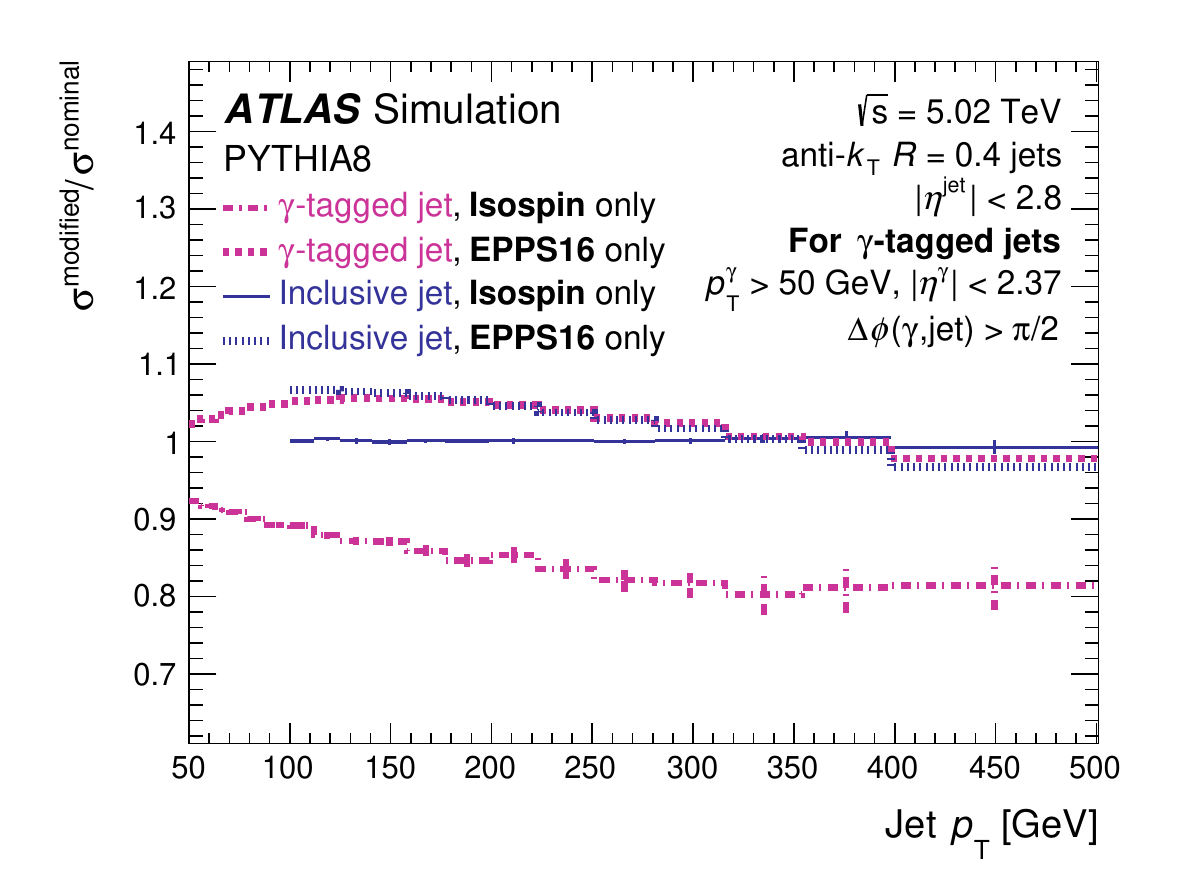}
\caption{
\PYTHIA8-based evaluation of the impact of the isospin and nPDF effects for the two jet samples, shown as the ratio of the modified cross-section to the nominal one in \ppp collisions. The isospin (nPDF) effect for inclusive jets is shown as a solid (dotted) line, and for photon-tagged jets as dot-dashed (dashed) line.}
\label{fig:Raa_isospin_nPDF_pythia}
\end{center}
\end{figure}
 
The possible quantitative impact of these effects can be explored using the generator-level simulation samples described at the end of Section~\ref{sec:simulation}. To determine the impact of the isospin and nPDF effects, the simulated cross-section in Pb+Pb events or in nPDF-weighted \ppp events, respectively, was compared with that in the original sample of \ppp events. The ratios of these modified cross-sections to the cross-section in \ppp collisions are shown in Figure~\ref{fig:Raa_isospin_nPDF_pythia} separately for photon-tagged and inclusive jets. While the isospin effect for inclusive jets is negligible, it causes the photon-tagged jet spectrum (and thus $R_\mathrm{AA}$) in Pb+Pb collisions to decrease by 10--20\% in the \ptj range of 100--300~\GeV. The isospin effect is stronger at larger \ptj as the parton in the nucleus involved in the parton--parton scattering is more likely to come from a valence (up/down) quark at large Bjorken-$x$ range.  The nPDF effects on the photon-tagged and inclusive jet \RAA are similar, leading to approximately a 5\% enhancement at 100~\GeV\ (an increase in the nuclear parton densities in the `anti-shadowing' region) that then decreases with increasing $p_\mathrm{T}$. Given the similar nPDF effects, it can be seen that the isospin effect for photon-tagged jets has the dominant impact in the comparisons. It decreases the photon-tagged jet yield in Pb+Pb events, thus causing an overestimate of the energy loss effects under the naive interpretation of $S_\mathrm{loss}$ and \RAA.
 
To test the potential impact of these effects on the $S_\mathrm{loss}$ results, the energy loss study is repeated after dividing the measured $R_\mathrm{AA}$ values by the simulation-derived values in Figure~\ref{fig:Raa_isospin_nPDF_pythia} to approximately correct for these effects. The updated $S_\mathrm{loss}$ values are shown as dashed lines in Figure~\ref{fig:eloss_incJet_phoJet}. It can be seen that the differences in energy loss between photon-tagged jets and inclusive jets becomes even larger after accounting for the isospin and nPDF effects, further strengthening the evidence that quark-initiated jets lose less energy than gluon-initiated ones.

\subsection{Theoretical comparisons}
 
The $R_\mathrm{AA}$ results are compared with theoretical calculations of jet energy loss in the QGP that model the colour-charge dependence of the parton-QGP interaction in various ways. As discussed above, it is important for such calculations to properly model details such as the photon production processes (i.e., including fragmentation photons), the spectral shape, and the impact of the isospin and nPDF for a consistent comparison with the data.
The five calculations described below typically meet most but not necessarily all these criteria.
 
The calculation from Takacs \textit{et al.}~\cite{Mehtar-Tani:2021fud, Takacs:2021bpv} includes a resummation of energy loss effects from hard, vacuum-like emissions occurring in the medium and the modelling of soft energy flow and recovery at the jet cone. The Takacs \textit{et al.} calculations are presented with a range of the jet-medium coupling parameter $g_\mathrm{med} = 2.2$--$2.3$. The predictions in Refs.~\cite{Ke:2018jem, Ke:2018tsh} are based on a linearised Boltzmann equation with diffusion model (LIDO). The LIDO calculations are presented with a range of values for the parameter $\mu = 1.3\pi{T}$--$1.8\pi{T}$, where $T$ is the medium temperature and $\mu$ controls the strength of the parton coupling to the medium. The predictions labelled $\mathrm{SCET_{G}}$ are perturbative calculations performed within the framework of soft-collinear effective field theory with Glauber gluons in the soft-gluon-emission (energy-loss) limit~\cite{Kang:2017xnc, Li:2018xuv, Li:2019dre}, with the width of the band in the Figures corresponding to the range of jet-medium coupling $g = 2.0 \pm 0.2$. The linear Boltzmann transport (LBT) model predictions~\cite{He:2018xjv} include elastic and inelastic processes based on perturbative QCD for both jet shower and recoil medium partons as they propagate through a QGP. JEWEL~\cite{KunnawalkamElayavalli:2016ttl} is a MC event generator that simulates QCD jet evolution in heavy-ion collisions, including radiative and elastic energy loss processes, and was configured without including medium recoils in the jet reconstruction, but with accounting for the isospin effect.
 
\begin{figure}[t!]
\begin{center}
\includegraphics[width=0.98\textwidth]{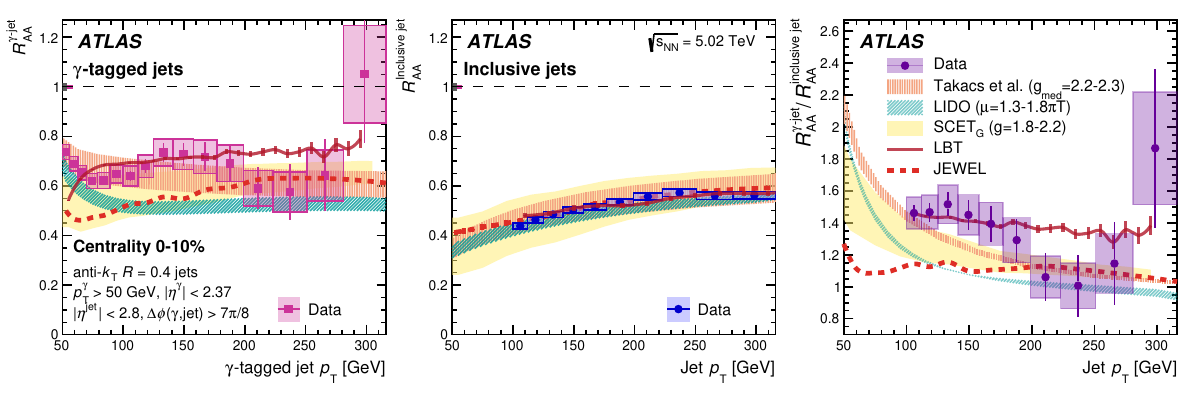}
\caption{
Comparison of \RAA between data and various theoretical predictions for (left panel) photon-tagged jets and  (middle panel) inclusive jets. The right panel shows \raagj/\raainc\ compared between data and theory predictions.
The vertical bars associated with symbols of the photon-tagged jet data indicate the statistical uncertainties and the total systematic uncertainties are shown as boxes in each \ptj bin. For inclusive jets, the boxes around the points indicate combined statistical and systematic uncertainties, although they are dominated by the latter. The bands of the theoretical calculations represent ranges of model parameters (see text). The vertical bars associated with the LBT calculation indicate the statistical uncertainties.
The shaded bars on the left of the axis at $\RAA = 1$ indicate the $p_\mathrm{T}$-independent uncertainties associated with the luminosity in \ppp collisions and \TAAavg for 0--10\% Pb+Pb collisions, respectively.
}
\label{fig:Raa_compTheory_only0to10}
\end{center}
\end{figure}
 
The left and middle panels of Figure~\ref{fig:Raa_compTheory_only0to10} show the \RAA of photon-tagged jets and inclusive jets, respectively, in 0--10\% central \pbpb collisions compared with the theoretical predictions. The ratio \raagj/\raainc is shown in the right panel, which in the theoretical predictions leads to the cancellation of some uncertainties common to both \RAA calculations. The inclusive jet \RAA, a commonly used benchmark to fix free parameters in theoretical models, is well described by all of the calculations. All the calculations except JEWEL qualitatively predict that the photon-tagged jet \RAA should be closer to unity than the inclusive jet \RAA, but the specific magnitude as a function of \ptj varies.  The photon-tagged jet \RAA data points are generally larger than the central values of many of the calculations, but they are compatible with the LBT model and with the calculations by Takacs \textit{et al.} and $\mathrm{SCET_{G}}$ within the range of their respective model parameters. Notably, several of the models predict the increase of the photon-tagged jet $R_\mathrm{AA}$ with decreasing \ptj observed in data at \ptj$\lesssim 80$~\GeV. The models further predict that the \raagj/\raainc ratio systematically decreases with increasing \ptj, as the quark-initiated fraction in the two samples become more similar, which is also qualitatively present in the data. However, the agreement with the models is worse, with only the LBT model describing the measured double ratio. Since these models otherwise described the inclusive jet $R_\mathrm{AA}$ well, this additional comparison highlights the need to test them against multiple observables simultaneously to evaluate the description of the colour-charge dependence of energy loss.

\FloatBarrier 
\section{Conclusion}
 
This Letter presents a measurement of photon-tagged jet production in 1.7~nb$^{-1}$ of Pb+Pb and 260~pb$^{-1}$ of \ppp\ collisions at \comHI with the ATLAS detector. The cross-section of jets produced opposite in azimuth ($\Delta\phi > 7\pi/8$) to a $\ptg > 50$~\GeV\ isolated photon is reported as a function of \ptj. This selection results in a sample of jets with a steeply falling \pt distribution and a large fraction of quark-initiated jets. The nuclear modification factor, \raa, for photon-tagged jets is found to be suppressed below unity in a way that varies with centrality but only weakly with \ptj in the measured range.
The fractional energy loss, $S_\mathrm{loss}$, is determined to be approximately 0.10 with no strong $p_\mathrm{T}$ dependence within uncertainties in the 0--10\% centrality interval.
The photon-tagged jet \raa ($S_\mathrm{loss}$)
is significantly higher (lower) than that for inclusive jets at the same \ptj and centrality, which instead have a large gluon-initiated jet fraction. The results are compared with a variety of theoretical calculations, which qualitatively describe aspects of the ordering between photon-tagged and inclusive jets, but tend to over-predict the amount of energy loss for the former. The data provide the strongest confirmation to date of larger jet quenching for gluon jets compared with quark jets.

\section*{Acknowledgements}


We thank CERN for the very successful operation of the LHC, as well as the
support staff from our institutions without whom ATLAS could not be
operated efficiently.
 
We acknowledge the support of
ANPCyT, Argentina;
YerPhI, Armenia;
ARC, Australia;
BMWFW and FWF, Austria;
ANAS, Azerbaijan;
CNPq and FAPESP, Brazil;
NSERC, NRC and CFI, Canada;
CERN;
ANID, Chile;
CAS, MOST and NSFC, China;
Minciencias, Colombia;
MEYS CR, Czech Republic;
DNRF and DNSRC, Denmark;
IN2P3-CNRS and CEA-DRF/IRFU, France;
SRNSFG, Georgia;
BMBF, HGF and MPG, Germany;
GSRI, Greece;
RGC and Hong Kong SAR, China;
ISF and Benoziyo Center, Israel;
INFN, Italy;
MEXT and JSPS, Japan;
CNRST, Morocco;
NWO, Netherlands;
RCN, Norway;
MEiN, Poland;
FCT, Portugal;
MNE/IFA, Romania;
MESTD, Serbia;
MSSR, Slovakia;
ARRS and MIZ\v{S}, Slovenia;
DSI/NRF, South Africa;
MICINN, Spain;
SRC and Wallenberg Foundation, Sweden;
SERI, SNSF and Cantons of Bern and Geneva, Switzerland;
MOST, Taiwan;
TENMAK, T\"urkiye;
STFC, United Kingdom;
DOE and NSF, United States of America.
In addition, individual groups and members have received support from
BCKDF, CANARIE, Compute Canada and CRC, Canada;
PRIMUS 21/SCI/017 and UNCE SCI/013, Czech Republic;
COST, ERC, ERDF, Horizon 2020 and Marie Sk{\l}odowska-Curie Actions, European Union;
Investissements d'Avenir Labex, Investissements d'Avenir Idex and ANR, France;
DFG and AvH Foundation, Germany;
Herakleitos, Thales and Aristeia programmes co-financed by EU-ESF and the Greek NSRF, Greece;
BSF-NSF and MINERVA, Israel;
Norwegian Financial Mechanism 2014-2021, Norway;
NCN and NAWA, Poland;
La Caixa Banking Foundation, CERCA Programme Generalitat de Catalunya and PROMETEO and GenT Programmes Generalitat Valenciana, Spain;
G\"{o}ran Gustafssons Stiftelse, Sweden;
The Royal Society and Leverhulme Trust, United Kingdom.
 
The crucial computing support from all WLCG partners is acknowledged gratefully, in particular from CERN, the ATLAS Tier-1 facilities at TRIUMF (Canada), NDGF (Denmark, Norway, Sweden), CC-IN2P3 (France), KIT/GridKA (Germany), INFN-CNAF (Italy), NL-T1 (Netherlands), PIC (Spain), ASGC (Taiwan), RAL (UK) and BNL (USA), the Tier-2 facilities worldwide and large non-WLCG resource providers. Major contributors of computing resources are listed in Ref.~\cite{ATL-SOFT-PUB-2021-003}.


\printbibliography
 
\clearpage
 
\begin{flushleft}
\hypersetup{urlcolor=black}
{\Large The ATLAS Collaboration}

\bigskip

\AtlasOrcid[0000-0002-6665-4934]{G.~Aad}$^\textrm{\scriptsize 102}$,
\AtlasOrcid[0000-0002-5888-2734]{B.~Abbott}$^\textrm{\scriptsize 120}$,
\AtlasOrcid[0000-0002-1002-1652]{K.~Abeling}$^\textrm{\scriptsize 55}$,
\AtlasOrcid[0000-0001-5763-2760]{N.J.~Abicht}$^\textrm{\scriptsize 49}$,
\AtlasOrcid[0000-0002-8496-9294]{S.H.~Abidi}$^\textrm{\scriptsize 29}$,
\AtlasOrcid[0000-0002-9987-2292]{A.~Aboulhorma}$^\textrm{\scriptsize 35e}$,
\AtlasOrcid[0000-0001-5329-6640]{H.~Abramowicz}$^\textrm{\scriptsize 151}$,
\AtlasOrcid[0000-0002-1599-2896]{H.~Abreu}$^\textrm{\scriptsize 150}$,
\AtlasOrcid[0000-0003-0403-3697]{Y.~Abulaiti}$^\textrm{\scriptsize 117}$,
\AtlasOrcid[0000-0003-0762-7204]{A.C.~Abusleme~Hoffman}$^\textrm{\scriptsize 137a}$
% ...
\bigskip
\\

% ...
$^{29}$Physics Department, Brookhaven National Laboratory, Upton NY; United States of America.\\
% ...
$^{35}$$^{(a)}$Facult\'e des Sciences Ain Chock, R\'eseau Universitaire de Physique des Hautes Energies - Universit\'e Hassan II, Casablanca;$^{(b)}$Facult\'{e} des Sciences, Universit\'{e} Ibn-Tofail, K\'{e}nitra;$^{(c)}$Facult\'e des Sciences Semlalia, Universit\'e Cadi Ayyad, LPHEA-Marrakech;$^{(d)}$LPMR, Facult\'e des Sciences, Universit\'e Mohamed Premier, Oujda;$^{(e)}$Facult\'e des sciences, Universit\'e Mohammed V, Rabat;$^{(f)}$Institute of Applied Physics, Mohammed VI Polytechnic University, Ben Guerir; Morocco.\\
% ...
$^{49}$Fakult\"{a}t Physik , Technische Universit{\"a}t Dortmund, Dortmund; Germany.\\
% ...
$^{55}$II. Physikalisches Institut, Georg-August-Universit\"{a}t G\"ottingen, G\"ottingen; Germany.\\
% ...
$^{102}$CPPM, Aix-Marseille Universit\'e, CNRS/IN2P3, Marseille; France.\\
% ...
$^{117}$Department of Physics, New York University, New York NY; United States of America.\\
% ...
$^{120}$Homer L. Dodge Department of Physics and Astronomy, University of Oklahoma, Norman OK; United States of America.\\
% ...
$^{137}$$^{(a)}$Departamento de F\'isica, Pontificia Universidad Cat\'olica de Chile, Santiago;$^{(b)}$Millennium Institute for Subatomic physics at high energy frontier (SAPHIR), Santiago;$^{(c)}$Instituto de Investigaci\'on Multidisciplinario en Ciencia y Tecnolog\'ia, y Departamento de F\'isica, Universidad de La Serena;$^{(d)}$Universidad Andres Bello, Department of Physics, Santiago;$^{(e)}$Instituto de Alta Investigaci\'on, Universidad de Tarapac\'a, Arica;$^{(f)}$Departamento de F\'isica, Universidad T\'ecnica Federico Santa Mar\'ia, Valpara\'iso; Chile.\\
% ...
$^{150}$Department of Physics, Technion, Israel Institute of Technology, Haifa; Israel.\\
$^{151}$Raymond and Beverly Sackler School of Physics and Astronomy, Tel Aviv University, Tel Aviv; Israel.\\
% ...
$^{a}$ Also Affiliated with an institute covered by a cooperation agreement with CERN.\\
$^{b}$ Also at An-Najah National University, Nablus; Palestine.\\
$^{c}$ Also at APC, Universit\'e Paris Cit\'e, CNRS/IN2P3, Paris; France.\\
$^{d}$ Also at Borough of Manhattan Community College, City University of New York, New York NY; United States of America.\\
$^{e}$ Also at Center for High Energy Physics, Peking University; China.\\
$^{f}$ Also at Center for Interdisciplinary Research and Innovation (CIRI-AUTH), Thessaloniki; Greece.\\
% ...

\end{flushleft}


\end{document}